\documentclass[prl, twocolumn, nofootinbib, showpacs, superscriptaddress]{revtex4} 
\usepackage{graphicx}  
\usepackage{dcolumn}   
\usepackage{bm}        
\usepackage{amssymb}   
\usepackage{epsfig}
\usepackage{amsmath} 

\begin{document}
\title{Toward a Universal Formulation of the Halo Mass Function}
\author{P.S. Corasaniti}
\author{I. Achitouv}
\address{Laboratoire Univers et Th\'eories (LUTh), UMR 8102 CNRS, Observatoire de Paris,
Universit\'e Paris Diderot, \\ 5 Place Jules Janssen, 92190 Meudon, France}
\date{\today}
\begin{abstract} 
We compute the dark matter halo mass function using the excursion set formalism for 
a diffusive barrier with linearly drifting average which captures the main 
features of the ellipsoidal collapse model. We evaluate the non-Markovian
corrections due to the sharp filtering of the linear density field in
real space with a path-integral method. We find an unprecedented
agreement with N-body simulation data with deviations $\lesssim 5\%$ over the
range of masses probed by the simulations. This indicates that the Excursion Set 
in combination with a realistic modelling of the collapse threshold can provide a 
robust estimation of the halo mass function.
\end{abstract}

\maketitle
A large body of evidence suggests that dark matter (DM)
plays a crucial role in the formation, evolution and spatial distribution
of cosmic structures \cite{Spergel2003,Tegmark2004,Clowe2006,Massey2007}. 
Central to the DM paradigm is the idea that initial density fluctuations 
grow under gravitational instability eventually collapsing into virialized objects, the halos. 
It is inside these gravitationally bounded structures that cooling baryonic gas falls in to form the stars 
and galaxies we observe today. Consequently, the study of the halo mass distribution
is of primary importance in cosmology. 
In the Press-Schechter approach
\cite{PressSchechter1974}, the number of halos in the mass range
$[M,M+dM]$ can be written as
\begin{equation}
\frac{dn}{dM}=f(\sigma)\frac{\bar{\rho}}{M^2}\frac{d\log{\sigma^{-1}}}{d\log{M}},\label{mf}
\end{equation}
where $\bar{\rho}$ is the background matter density and $\sigma(M)$ is the root-mean-square
fluctuation of the linear dark matter density field smoothed on a scale $R(M)$ (containing a mass $M$), with
\begin{equation}
\sigma^2(M)\equiv S(M)=\frac{1}{2\pi^2}\int dk~k^2P(k)\tilde{W}^2[k,R(M)],
\end{equation}
where $P(k)$ is the linear DM power spectrum and $\tilde{W}(k,R)$ is
the Fourier transform of the smoothing (filter) function
in real space. In Eq.~(\ref{mf}), the function $f(\sigma)=2\sigma^2\mathcal{F}(\sigma^2)$,
known as `multiplicity function', encodes the effects of the gravitational processes
responsible for the formation of halos through its dependence on
$\mathcal{F}(S)\equiv dF/dS$, with $F(S)$ being the fraction of mass elements
in halos of mass $>M(S)$. Hereafter, we will refer to $f(\sigma)$ 
simply as the halo mass function.

The collapse of halos is a highly nonlinear gravitational process that has been
primarily investigated using numerical N-body simulations. Over the past few years several numerical studies 
have measured $f(\sigma)$ at few percent uncertainty level for various 
cosmologies and using different halo detection algorithms (see e.g. \cite{Tinker2008,Courtin2011,Crocce2010,Suman2011}). 
On the other hand, we still lack an accurate theoretical estimation of the halo mass function.
Following the seminal work by Press and Schechter \cite{PressSchechter1974},  
the excursion set theory \cite{Bond1991} has provided us with a consistent
mathematical framework for computing $f(\sigma)$ 
from the statistical properties of the initial DM density field
(for a review see \cite{Zentner2007}). Nevertheless, an analytical derivation of $f(\sigma)$
can be obtained only for a top-hat filter in Fourier
space (sharp-k filter). Although Monte-Carlo simulations can be used
in the case of generic filters (see e.g. \cite{Bond1991,Percival2001}), most
of the work in the literature has focused on the modeling of the halo
collapse conditions and the comparison with N-body simulations, while assuming
the sharp-$k$ filter (see, e.g., \cite{Sheth1998,Sheth2001,Shen2006,JunHui2006}).
However, such a smoothing function does not correspond to any realistic halo mass
definition. The issue has been recently addressed by Maggiore and
Riotto \cite{MaggioreRiotto2010a} who made a major contribution by introducing a path-integral method that extends
the analytical computation to generic filters. 

In this Letter we present the first thorough comparison
against N-body simulation data of the excursion set mass function 
with top-hat filter in real space for
a stochastic barrier model which encapsulates the
main characteristics of the ellipsoidal collapse of dark matter.
A detailed derivation of these results is given in a
companion paper \cite{PSCIA2011}. 

Let us consider the DM density contrast, $\delta({\bf x})$, smoothed on the scale $R$, 
\begin{equation}
\delta({\bf x},R)=\int d^3y\, W(|{\bf x}-{\bf y}|,R)\,\delta({\bf y}),
\end{equation} 
where $W({\bf x},R)$ is the smoothing function in real space. Bond et al.~\cite{Bond1991} have shown
that at any given point in space, $\delta({\bf x},R)$ performs a
random walk as a function of the variance of the
smoothed linear density field $S(R)$. The formation of halos of mass $M$ corresponds to trajectories 
$\delta(S)$ crossing for the first time a barrier $B$ at $S(M)$,
i.e., $\delta(S)=B$, where the value of $B$ depends on the assumed gravitational collapse criterion. 
In the case of the spherical collapse model \cite{GunnGott1972} $B=\delta_c$, 
that is the linearly extrapolated density of a top-hat spherical perturbation at the time of collapse. 
Then, the evaluation of $f(\sigma)$ is reduced to computing the rate at which 
the random walks hit the barrier for the first time, i.e., $\mathcal{F}(S)=dF/dS$.
 
The nature of the random walk depends on the filtering procedure,
which specifies the relation between the smoothing scale $R$ and the halo mass definition $M$. 
For a sharp-$k$ filter, $\tilde{W}(k,R)=\theta(1/R-k)$, and Gaussian initial conditions, 
$\delta(S)$ performs a Markov random walk described by the Langevin equation:
\begin{equation}
\frac{\partial \delta}{\partial S}=\eta_\delta(S),\label{dd}
\end{equation}
with noise $\eta_\delta(S)$ such that $\langle\eta_\delta(S)\rangle=0$ and 
$\langle\eta_\delta(S)\eta_\delta(S')\rangle=\delta_D(S-S')$, where $\delta_D$ is the Dirac-function 
(for the full derivation, see,
e.g., \cite{Zentner2007,MaggioreRiotto2010a}). As first shown in
\cite{Bond1991}, the probability distribution of the trajectories
satisfies a simple Fokker-Planck equation with absorbing boundary at $\delta(S)=\delta_c$.  
The resulting first-crossing distribution gives the Press-Schechter 
formula \cite{PressSchechter1974} with the correct normalization factor (the so called `extended Press-Schechter').

However, the spherical collapse model is a simplistic approximation 
of the nonlinear evolution of matter density fluctuations. As shown in
\cite{Doroshkevich1970}, initial Gaussian perturbations are highly nonspherical. 
Hence, the collapse of a homogeneous ellipsoid (see, e.g., \cite{EisensteinLoeb1995}) should 
provide a far better description. In such a model the critical density
threshold depends on the eigenvalues of the deformation tensor, which are random variables with
probability distributions that depend on the statistics of the linear
density field \cite{Doroshkevich1970,Bardeenetal1986,Monaco1995,Audit1997,Lee1998,Sheth2001,Desjacques2008}.
Because of this, the barrier behaves as a stochastic variable itself, 
performing a random walk whose properties depend on the specificities
of the collapse model considered. For example, Sheth et
al. \cite{Sheth2001} showed that the average of the barrier is $\langle B(S)\rangle=\delta_c[1+\beta(S/\delta_c^2)^{\gamma}]$, with
$\beta=0.47$ and $\gamma=0.615$. 

The recent analysis of halos in N-body simulations has confirmed the stochastic barrier
hypothesis \cite{Robertson2009}. Maggiore and Riotto
\cite{MaggioreRiotto2010b} have modeled these features assuming a
stochastic barrier with average $\langle B(S)\rangle=\delta_c$ and variance 
$\langle(B-\langle B(S)\rangle)^2\rangle=S\,D_B$, where $D_B$ is a
constant diffusion coefficient. 
Here, we improve their barrier model by assuming a Gaussian diffusion with 
linearly drifting average $\langle B(S)\rangle=\delta_c+\beta S$
\cite{Sheth1998} which approximates the ellipsoidal collapse
prediction \cite{Sheth2001}. Recently, a general analysis of 
nondiffusive moving barriers has been presented in \cite{DeSimone2010}.
However, this work has mainly focused on the mass function 
in the presence of Non-Gaussian initial conditions rather than the comparison 
with Gaussian N-body simulations. 
The Langevin equation for this barrier model reads as
\begin{equation}
\frac{\partial B}{\partial S}=\beta+\eta_B(S),\label{db}
\end{equation}
where the noise $\eta_B(S)$ is characterized by $\langle\eta_B(S)\rangle=0$ and 
$\langle\eta_B(S)\eta_B(S')\rangle=D_B\,\delta_D(S-S')$. Without loss
of generality we can assume that $\eta_B(S)$ and $\eta_\delta(S)$ are uncorrelated.
It is convenient to introduce $Y=B-\delta$ and rewrite Eqs.~(\ref{dd}) and (\ref{db}) as a single Langevin equation:
\begin{equation}
\frac{\partial Y}{\partial S}=\beta+\eta(S),\label{dy}
\end{equation}
with white noise $\eta(S)=\eta_\delta(S)+\eta_B(S)$ such that $\langle \eta(S)\rangle=0$ and
$\langle \eta(S)\eta(S')\rangle=(1+D_B)\delta(S-S')$. The
Fokker-Planck equation associated with Eq.~(\ref{dy}) and describing the
probability $\Pi_{0}(Y_0,Y,S)$ reads as
\begin{equation}
\frac{\partial \Pi_{0}}{\partial S}=-\beta\frac{\partial \Pi_{0}}{\partial Y}+\frac{1+D_B}{2}\frac{\partial^2\Pi_{0}}{\partial Y^2},\label{FP}
\end{equation}
where we indicate with the ``$0$'' underscore the fact that $\Pi_{0}$ is associated to a Markov process. 

The system starts at $\{\delta(0)=0,B(0)=\delta_c\}$; hence, we solve Eq.~(\ref{FP}) 
with initial condition $Y_0=\delta_c$ and impose the absorbing
boundary condition at $Y=0$, i.e., $\Pi_{0}(0,S)=0$. For a concise notation
we omit the dependence on $Y_0$ and simply refer to $\Pi_{0}(Y,S)$. 
By rescaling the variable $Y\rightarrow \tilde{Y}=Y/\sqrt{1+D_B}$, 
a factorizable solution can be found in the form $\Pi_{0}(\tilde{Y},S)=U(\tilde{Y},S)\exp[c(\tilde{Y}-cS/2)]$, where 
$c=\beta/\sqrt{1+D_B}$ and $U(\tilde{Y},S)$ satisfies a diffusion equation. Using the above initial condition, 
the latter can be solved with the image method \cite{Redner2001} or by Fourier transform. Thus, we obtain
\begin{equation}
\Pi_{0}(Y,S)=\frac{e^{\frac{\beta}{1+D_B}(Y-Y_0-\beta\frac{S}{2})}}{\sqrt{2\pi S(1+D_B)}}\left[e^{-\frac{(Y-Y_0)^2}{2S(1+D_B)}}-e^{-\frac{(Y+Y_0)^2}{2S(1+D_B)}}\right].\label{pimarkov}
\end{equation}
In general the Fokker-Planck equation for random walks with nonlinear
biased diffusion and absorbing boundary condition does not have an
exact analytic solution. This is why we have assumed the linearly
drifting average barrier rather than the prediction of the ellipsoidal
collapse model \cite{Sheth2001}. As we will see later, having an exact
analytical solution greatly simplify the evaluation of the corrections due to the smoothing function.
We should remark that the above solution is defined only for $Y>0$. Since the number
of trajectories is conserved, then the first-crossing distribution is
obtained by deriving $\int_0^S\mathcal{F}_{0}(S')dS'=1-\int_0^\infty\Pi_0(Y,S)dY$ from which we
finally obtain the Markovian mass function 
\begin{equation}\label{flin}
f_{0}(\sigma)=\frac{\delta_c}{\sigma\sqrt{1+D_B}}\sqrt{\frac{2}{\pi}}e^{-\frac{(\delta_c+\beta \sigma^2)^2}{2 \sigma^2(1+D_B)}},
\end{equation}
for $\beta=0$ and $D_B=0$ this coincides with the standard Markovian solution that gives the extended Press-Schechter formula, while for
$D_B=0$ we recover the solution for the nondiffusive linearly drifting barrier \cite{Zentner2007}. 

As mentioned earlier, a crucial point of this derivation is the assumption of the sharp-$k$ filter. 
In numerical N-body simulations the mass definition depends on the halo detection algorithm. 
For instance, the spherical overdensity (SOD) halo finder detects
halos as groups of particles in a spherical regions of radius
$R_\Delta$ containing a density $\rho_\Delta=\Delta\bar{\rho}$, with $\Delta$
an overdensity parameter usually fixed to $\Delta=200$. Thus, the halo
mass is $M=4/3\pi R_\Delta^3\rho_\Delta$, which is equivalent
to having a sharp-x filter, or
$\tilde{W}(k,R)=(3/kR)[\sin(kR)-(kR)\cos(kR)]$. 
However, in this case the stochastic evolution of the
system is no longer Markovian. Hence, in order to consistently compare
the excursion set mass function with SOD estimates of $f(\sigma)$ it
is necessary to account for the correlations induced by $\tilde{W}(k,R)$. 

Maggiore and Riotto \cite{MaggioreRiotto2010a} have shown that these
correlations can be treated as perturbations about the ``zero''-order 
Markovian solution. More specifically, the noise variable $\eta(S)$ acquires a perturbative correction, 
$\langle\eta(S)\eta(S')\rangle=(1+D_B)\delta_D(S-S')+\Delta(S,S')$, which in the case of the sharp-x filter
can be approximated by $\Delta(S,S')\approx\kappa S(S'-S)/S'$. For the
concordance $\Lambda$ Cold DM model we find $\kappa\approx 0.47$.
Using the path-integral technique described in
\cite{MaggioreRiotto2010a}, we compute the corrections 
to $\Pi_{0}(Y,S)$ to first order in $\kappa$. These consist of a ``memory'' term,
\begin{equation}
\Pi_{1}^{m}=-{\partial_Y}\int_0^SdS'\Delta(S',S)\Pi^{f}_{0}(Y_0,0,S')\Pi^{f}_{0}(0,Y,S-S'),\label{mem}
\end{equation}
and a ``memory-of-memory'' term
\begin{eqnarray}
&\Pi_{1}^{m-m}=\int_0^SdS'\int_{S''}^SdS''\Delta(S',S'')\Pi^{f}_{0}(Y_0,0,S')\times&\nonumber\\
&\times\Pi^{f}_{0}(0,0,S''-S')\Pi^{f}_{0}(0,Y,S-S'),&\label{memmem}
\end{eqnarray}
where $\Pi^{f}_{0}(Y_0,0,S)$, $\Pi^{f}_{0}(0,Y,S)$ and $\Pi^{f}_{0}(0,0,S)$ in Eqs.~(\ref{mem}) and (\ref{memmem}) 
are given by the finite time corrections of the Markovian solution near the barrier (see \cite{PSCIA2011}). 
We find
\begin{eqnarray}
\Pi_{0}^f(Y_0,0,S)&=&\frac{a\,Y_0}{S^{3/2}\sqrt{\pi}}e^{-\frac{a(Y_0+\beta S)^2}{2S}},\label{p1d}\\
\Pi_{0}^f(0,Y,S)&=&\frac{a\,Y}{S^{3/2}\sqrt{\pi}}e^{-\frac{a(Y-\beta S)^2}{2S}},\label{p2d}\\
\Pi_{0}^f(0,0,S)&=&\frac{1}{S^{3/2}}\sqrt{\frac{a}{2\pi}},
\end{eqnarray}
where $a\equiv 1/(1+D_B)$. 
Eq.~(\ref{mem}) can be computed analytically, we find
\begin{equation}
\Pi_{1}^m=-\tilde{\kappa}\,a\,Y_0\,{\partial_Y}\left\{Y e^{a\beta\left(Y-Y_0-\beta\frac{S}{2}\right)}\,{\rm Erfc}\left[\sqrt{\frac{a}{2S}}(Y_0+Y)\right]\right\},\label{pi1m}
\end{equation}
where $\tilde{\kappa}=\kappa/(1+D_B)$. Since Equation~(\ref{pi1m}) is linear in $Y$, the integration of 
$\mathcal{F}^m_{1}(S)=-\partial/\partial{S}\int_0^\infty\Pi_{1}^m\,dY$
vanishes. Thus, the memory term does not contribute to the mass
function independently of the barrier behavior (in agreement with
\cite{MaggioreRiotto2010a}).
\begin{figure}[t]
\includegraphics[scale=0.45]{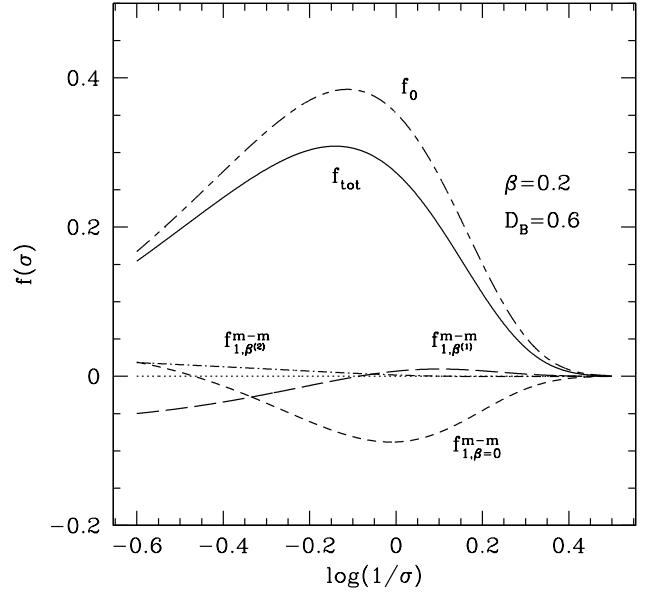}
\caption{Contributions to the halo mass function $f_\textrm{tot}$
  (solid line) for $\beta=0.2$ and $D_B=0.6$. The different curves correspond to the
Markovian mass function $f_0$ (dotted line), $f^{\textrm{m-m}}_{1,\beta=0}$
(short-dashed line), $f^\textrm{m-m}_{1,\beta^{(1)}}$ (long-dashed line),
$f^{\textrm{m-m}}_{1,\beta^{(2)}}$ (dot-short dashed line),
$f^{\textrm{m-m}}_{1,\beta^{(3)}}$ (dot-long dashed line).}
\label{fig1}
\end{figure}
The double integral in the memory-of-memory term cannot be computed
analytically, in such a case we expand the integrands in powers of
$\beta$ (given that from the ellipsoidal collapse we expect
$\beta<1$). By computing
$\mathcal{F}^{m-m}_{1}(S)=-\partial/\partial{S}\int_0^\infty\Pi_{1}^{m-m}dY$
and expressing the results directly in terms of $f(\sigma)$, we find
the non-Markovian correction to zero order in $\beta$ (i.e. $\beta=0$)
to be
\begin{equation}\label{beta0}
f_{(1),\beta=0}^{m-m}(\sigma)=-\tilde{\kappa}\frac{\delta_c}{\sigma}\sqrt{\frac{2a}{\pi}}\left[e^{-\frac{a\delta_c^2}{2\sigma^2}}-\frac{1}{2}\Gamma\left(0,\frac{a\delta_c^2}{2\sigma^2}\right)\right],
\end{equation}
where $\Gamma(0,z)$ is the incomplete Gamma function. Not surprisingly this expression coincides with 
the memory-of-memory term in \cite{MaggioreRiotto2010a}. The first order correction in $\beta$ is given by
\begin{equation}\label{beta1}
f_{1,\beta^{(1)}}^{m-m}(\sigma)=-\beta\,a\,\delta_c\left[f_{1,\beta=0}^{m-m}(\sigma)+\tilde{\kappa}\,\textrm{Erfc}\left(\frac{\delta_c}{\sigma}\sqrt{\frac{a}{2}}\right)\right],
\end{equation}
and the second order reads 
\begin{equation}\label{beta2}
\begin{split}
&f_{1,\beta^{(2)}}^{m-m}(\sigma)=\beta^2\,a\,\delta_c\,\tilde{\kappa}\biggl\{a\,\delta_c\,\textrm{Erfc}\left(\frac{\delta_c}{\sigma}\sqrt{\frac{a}{2}}\right)+\\
&+\sigma\sqrt{\frac{a}{2\pi}}\biggl[e^{-\frac{a\delta_c^2}{2\sigma^2}}\left(\frac{1}{2}-\frac{a\delta_c^2}{\sigma^2}\right)+\frac{3}{4}\frac{a\delta_c^2}{\sigma^2}\Gamma\left(0,\frac{a\delta_c^2}{2\sigma^2}\right)\biggr]\biggr\}.
\end{split}
\end{equation}
For $\beta/(1+D_B)<1$, corrections $\mathcal{O}(>\beta^2)$ are
negligible (see, e.g., Fig.~\ref{fig1}), hence, Eqs.~(\ref{flin}) and 
(\ref{beta0})-(\ref{beta2}) give the relevant contributions to
the mass function.
\begin{figure}[t]
\includegraphics[scale=0.45]{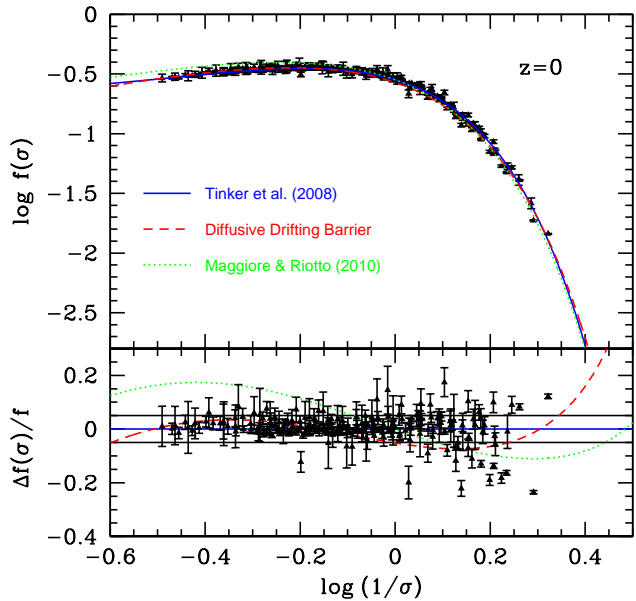}
\caption{(Upper panel) Halo mass function at $z=0$ given by the Tinker et al. fitting formula for $\Delta=200$ (solid blue line), 
diffusing drifting barrier with ${\beta}=0.057$ and ${D}_b=0.294$ (red dashed line) and Maggiore \& Riotto 
\cite{MaggioreRiotto2010b} with $D_B=0.235$ (green dotted line). Data points are from \cite{Tinker2008}. (Lower panel)
Relative difference with respect to the Tinker et al. fitting formula. The thin black solid lines indicates $5\%$ deviations.}
\label{fig2}
\end{figure}

In principle the values of $\beta$ and $D_B$ as well as their redshift
and cosmology dependence can be predicted in a given halo collapse model 
by computing the average and variance of the probability distribution
of the collapse density threshold. However, this requires a dedicated 
study which should also include environmental effects that have been shown to play an 
important role in determining the properties of the halo mass distribution \cite{Desjacques2008}.
This goes beyond the scope of this Letter.

Here, we take a different approach. $\beta$ and $D_B$ are
physical motivated model parameters which we can
calibrate against N-body simulation data, and test whether the
mass function derived above provides an acceptable description of the data. 
To this purpose we use the measurements of the halo mass function
obtained by Tinker et al. \cite{Tinker2008} 
using SOD(200) on a set of WMAP-1 yr and WMAP-3 yr cosmological N-body simulations. 
For these cosmological models the spherical collapse predicts $\delta_c=1.673$ at $z=0$ 
(for a detailed calculation see \cite{Courtin2011}). Using such a value, we run a likelihood Markov chain 
Monte Carlo analysis to confront the mass function previously computed 
against the data at $z=0$ in the prior parameter space $\log{\beta}=[-4,0]$ and $\log{D_B}=[-3,0]$. 
We find the best fit values to be ${\beta}=0.057$ and ${D}_b=0.294$. The data strongly constrain these parameters, 
with errors $\sigma_{\beta}=0.001$ and $\sigma_{D_B}=0.001$ respectively. 
In Fig.~\ref{fig2} (upper panel) we plot the corresponding mass function (red dash line) against the simulation data 
together with the four-parameter fitting formula by Tinker et al. \cite{Tinker2008} for $\Delta=200$ (solid blue line). 
For comparison we also plot the diffusive barrier case
by Maggiore and Riotto \cite{MaggioreRiotto2010b} which best fit the data with $D_B=0.235$ (green dotted line). 
In Fig.~\ref{fig2} (lower panel) we plot the relative differences with respect to the Tinker et al. formula.
We may notice the remarkable agreement of the diffusive drifting barrier with the data. Deviations with 
respect to Tinker et al. (2008) are $\lesssim 5\%$ level over the range of masses probed by the simulations. This
is quite impressive given the fact that our model depends only
on two physically motivated parameters.

In the upcoming years a variety of astrophysical observations will directly
probe $dn/dM$. The halo mass function we have derived here can provide the base for a through cosmological
model comparison. In a companion paper we will describe in detail the derivation 
of these results, as well as extensive
discussion on the redshift evolution of the mass function and halo bias. 

\begin{acknowledgments}
We are especially thankful to J. Tinker for kindly providing us with the mass function data. It is a pleasure to thank J.-M. Alimi, 
Y. Rasera, T. Riotto and R. Sheth for useful discussions. I. Achitouv is supported by the 
``Minist\`ere de l'Education Nationale, de la Recherche et de la Technologie'' (MENRT). 

\end{acknowledgments}


\begin{thebibliography}{99}  
\bibitem{Spergel2003} D. N. Spergel {\sl et al.},
  Astrophys. J. Suppl. Ser. {\bf 148}, 175 (2003). 
\bibitem{Tegmark2004} M. Tegmark {\sl et al.}, Phys. Rev. D. {\bf 69}, 103501 (2004).
\bibitem{Clowe2006} D. Clowe {\sl et al.}, Astrophys. J. {\bf 648}, L109 (2006).
\bibitem{Massey2007} R. Massey {\sl et al.}, Nature (London) {\bf 445}, 286
  (2007).
\bibitem{PressSchechter1974} W. H. Press and P. Schechter, Astrophys. J. {\bf 187}, 425 (1974).
\bibitem{Tinker2008} J. Tinker {\sl et al.}, Astrophys. J. {\bf 688}, 709 (2008).
\bibitem{Crocce2010} M. Crocce {\sl et al.}, Mont. Not. R. Astron. Soc. {\bf 403}, 1353 (2010).
\bibitem{Courtin2011} J. Courtin {\sl et al.},
  Mont. Not. R. Astron. Soc. \textbf{410}, 1911 (2011).
\bibitem{Suman2011} S. Batthacharya {\sl et al.},
  Astrophys. J. \textbf{732}, 122 (2011), arXiv:1005.2239.
\bibitem{Bond1991} J. R. Bond, S. Cole, G. Efstathiou and G. Kaiser, Astrophys. J. {\bf 379}, 440 (1991).
\bibitem{Zentner2007} A. R. Zentner, Int. J. Mod. Phys. D {\bf 16},
  763 (2007).
\bibitem{Percival2001} W.J. Percival, Mont. Not. R. Astron. Soc. {\bf
  327}, 1313 (2001). 
\bibitem{Sheth1998} R. K. Sheth, Mont. Not. R. Astron. {\bf 300}, 1057 (1998). 
\bibitem{Sheth2001} R. K. Sheth, H. J. Mo and G. Tormen,
  Mont. Not. R. Astron. {\bf 323}, 1 (2001).
\bibitem{Shen2006} J. Shen, T. Abel, H. J. Mo and R. K. Sheth,
  Astrophys. J. \textbf{645}, 783 (2006).
\bibitem{JunHui2006} J. Zhang and L. Hui, Astrophys. J. \textbf{641},
  641 (2006).
\bibitem{MaggioreRiotto2010a} M. Maggiore and A. Riotto, Astrophys. J. {\bf 711}, 907 (2010).
\bibitem{PSCIA2011} P.S. Corasaniti and I. Achitouv, PRD in press. 
\bibitem{GunnGott1972} J. E. Gunn and J. R. Gott III, Astrophys. J. {\bf 176}, 1 (1972).
\bibitem{Doroshkevich1970} A. G. Doroshkevich, Astrophyzika {\bf 3}, 175 (1970).
\bibitem{EisensteinLoeb1995} D. J. Eisenstein and A. Loeb,
  Astrophys. J. {\bf 439}, 520 (1995).
\bibitem{Bardeenetal1986} J. M. Bardeen, J. R. Bond, N. Kaiser and
  A. Szalay, Astrophys. J. \textbf{304}, 15 (1986).
\bibitem{Monaco1995} P. Monaco, Astrophys. J. {\bf 447}, 23 (1995).
\bibitem{Audit1997} E. Audit, R. Teyssier and J.-M. Alimi, Astron. and Astrophys. {\bf 325}, 439 (1997).
\bibitem{Lee1998} J. Lee and S. Shandarin, Astrophys. J. {\bf 500}, 14 (1998).
\bibitem{Desjacques2008} V. Desjacques, Mont. Not. R. Astron. {\bf 388}, 638 (2008).
\bibitem{Robertson2009} B. Robertson, A. Kravtsov, J. Tinker and A. Zentner, Astrophys. J. {\bf 696}, 636 (2009).
\bibitem{MaggioreRiotto2010b} M. Maggiore and A. Riotto, Astrophys. J. {\bf 717}, 515 (2010).
\bibitem{DeSimone2010} A. De Simone, M. Maggiore and A. Riotto, arXiv:1007.1903.
\bibitem{Redner2001} S. Redner, ``A guide to first-passage
  processes'', Cambridge University Press, Cambridge, U.K. (2001).
\end{thebibliography}
\end{document}